\documentclass[aps,prd,preprint,groupedaddress,showpacs,showkeys]{revtex4-1}

\pdfoutput=1

\usepackage[T1]{fontenc} 

\usepackage{epsfig}
\usepackage{graphicx}
\usepackage{amsmath}
\usepackage{amsfonts,amssymb}
\usepackage{color}

  \usepackage{slashed}
  \usepackage{url}
  \usepackage{amsthm}
\usepackage{float}
\usepackage{bm}
\usepackage{verbatim}

\begin{document}


\title{Nonlocal Nambu-Jona-Lasinio model and chiral chemical potential}


\author{Marco Frasca}
\email[]{marcofrasca@mclink.it}
\affiliation{Via Erasmo Gattamelata, 3 \\ 00176 Roma (Italy)}


\date{\today}

\begin{abstract}
We derive the critical temperature in a nonlocal Nambu-Jona-Lasinio model with the presence of a chiral chemical potential. The model we consider uses a form factor derived from recent studies of the gluon propagator in Yang-Mills theory and has the property to fit in excellent way the form factor arising from the instanton liquid picture for the vacuum of the theory. Nambu-Jona-Lasinio model is derived form quantum chromodynamics providing all the constants of the theory without any need for fits. We show that the critical temperature in this case always exists and increases as the square of the chiral chemical potential. The expression we obtain for the critical temperature depends on the mass gap that naturally arises from Yang-Mills theory at low-energy as also confirmed by lattice computations.
\end{abstract}






\maketitle


\section{Introduction}

Recent studies on lattice show that Yang-Mills theories develop a mass gap in the low energy limit. This is seen both in the spectrum \cite{Lucini:2004my,Chen:2005mg} and for the gluon propagator \cite{Bogolubsky:2007ud,Cucchieri:2007md,Oliveira:2007px}. On the theoretical side, several proposals have been put forward \cite{Cornwall:1981zr,Cornwall:2010bk,Dudal:2008sp,Frasca:2007uz,Frasca:2009yp} but none of them reached the status of a rigorous proof. Notwithstanding this difficulty, this fundamental result can be used to understand quantum chromodynamics (QCD) in the infrared limit. A very good approximation for the gluon propagator in the Landau gauge at lower energies is a free massive propagator as can be deduced from aforementioned references.

Existence of a mass gap and an analytical equation for the gluon propagator in a given gauge opens up the possibility to perform computations at low energies in QCD both at zero and finite temperature. We were able to prove in this way that a non-local Nambu-Jona-Lasinio (nNJL) model describes the low energy phenomenology of hadron physics \cite{Frasca:2008zp,Frasca:2013kka,Frasca:2012iv,Frasca:2012eq,Frasca:2011bd}. In Ref.\cite{Frasca:2011bd} we obtained the critical temperature at zero chemical potential for the chiral transition. This turns out in close agreement with lattice data \cite{Lucini:2012gg} and with preceding theoretical computations \cite{GomezDumm:2004sr}. 
A non-local Nambu-Jona-Lasinio model is capable of describing both chiral and deconfinement transitions, but without the cutoff of the local model. The model appears regularized in such a way to preserve anomalies, charges properly quantized and currents conserved. No extra cutoffs are needs as the effective interaction is finite at all orders \cite{Hell:2008cc}. Such a model, properly derived from QCD as we will show, will be used in this paper.

Our aim is to analyze a problem arisen when unbalanced chiral quark matter is present in a quark condensate. The question is how critical temperature changes due to the presence of chiral matter. This question was faced in \cite{Ruggieri:2011xc} to identify the critical end point of QCD. The idea is, at the thermodynamic equilibrium, to couple the chiral chemical potential, $\mu_5$,
to a chiral density quark operator, as also happens for the quark number density $\bar\psi\gamma_0\psi$ to the conjugated quark chemical potential $\mu$ (see \cite{Gatto:2011wc,Fukushima:2010fe,Chernodub:2011fr,Ruggieri:2011xc,Yu:2015hym,Yu:2014xoa,
Braguta:2015owi,Braguta:2015zta,Braguta:2016aov,Hanada:2011jb,Xu:2015vna} and references therein). 
%
Recent theoretical studies support the idea that the critical temperature should decrease with chiral chemical potential \cite{Gatto:2011wc,Fukushima:2010fe,Chernodub:2011fr,Ruggieri:2011xc,Yu:2015hym,Yu:2014xoa}. Recent lattice data have shown that critical temperature increases with $\mu_5$ \cite{Braguta:2015owi,Braguta:2015zta}.
This behavior of $T_c(\mu_5)$ was  predicted for the first time by universality arguments in~\cite{Hanada:2011jb}
and it has also been found later by solving Schwinger-Dyson equations at finite $\mu_5$~\cite{Xu:2015vna}. Recently, Ruggieri and Peng \cite{Ruggieri:2016cbq} draw this conclusion with a quark-meson model. We will strongly support their conclusions.

With our approach, we will show that a nNJL model provides a critical temperature increasing with the chiral chemical potential. We will show that, with the results provided in literature for the gluon propagator, the mass gap equation obtain always a solution both with $\mu_5=0$ and $\mu_5\ne 0$ and, in the latter case, it will depend on the square of it increasing the temperature. 

The paper is so structured. In Sec.~\ref{sec01} we show how NJL model can be seen as the low-energy limit of increasingly extended theories like QCD. In Sec.~\ref{sec1}, we complete the proof of the preceding section by presenting the 1- and 2-point functions and their agreement with lattice data. In Sec.~\ref{sec10}, we discuss the non-local NJL model obtained from QCD. In Sec.~\ref{sec2} we derive the critical temperature dependent on the chiral potential. Finally, in Sec.~\ref{sec3} conclusions are given.

\section{Derivation of the NJL model}
\label{sec01}

In this section, we will derive the Nambu-Jona-Lasinio model showing how increasingly complex bosonic-fermionic theories have it as effective theory when the bosonic degrees of freedom are integrated out, producing a quartic fermionic self-interaction. The result is exact for a Yukawa model but approximated when the bosonic degrees of freedom are self-interacting as for $\phi^4$ and Yang-Mills theories. We will always assume an evaluation of the generating functional
\begin{equation}
W[j,\bar\eta,\eta]=\int[d\mu]e^{i\int d^4xL+i\int d^4x\phi j+i\int d^4x[\bar\eta\psi+\bar\psi\eta]} 
\end{equation}
for generic bosonic $\phi$ and fermionic $\psi$ degrees of freedom and given a generic measure of integration for the path integral $[d\mu]$. $j$, $\bar\eta$, $\eta$ represents arbitrary currents introduced to derive the n-point functions.

\subsection{Yukawa model}

A free bosonic field, when coupled to a set of fermionic fields by Yukawa couplings, is completely equivalent to a non-local Nambu-Jona-Lasinio model.

We note that the bosonic field could have whatever spin but we limit the proof to the case of spin-0 extending to spin-1 when we analyze the case of QCD. So, let us consider the following Lagrangian
\begin{equation}
L=\frac{1}{2}\left[(\partial\phi)^2-m^2\phi^2\right]+\sum_f\bar\psi_f\left(i\slashed\partial-g\phi-m_f\right)\psi_f+j\phi+\sum_f\bar\eta_f\psi_f+\bar\psi_f\eta_f.
\end{equation}
We can integrate out the bosonic degree of freedom obtaining
\begin{equation}
L_f=\sum_f\bar\psi_f\left(i\slashed\partial-g^2\int d^4x'\Delta(x-x')\sum_f\bar\psi_f(x')\psi_f(x')-m_f\right)\psi_f
\end{equation}
provided
\begin{equation}
(\partial^2+m^2)\Delta(x-x')=\delta^4(x-x')
\end{equation}
the equation for the free propagator of the scalar field. 
We recognize this Lagrangian as that of a non-local Nambu-Jona-Lasinio model. In fact, the propagator of the scalar field in momenta space can be written down as
\begin{equation}
\Delta(p)=\frac{1}{p^2-m^2+i\epsilon}.
\end{equation}
Therefore, in the low-energy limit one gets
\begin{equation}
\Delta(x-x')=-\frac{1}{m^2}\delta^4(x-x')
\end{equation}
that yields the standard local NJL model provided we make the identification $G=g^2/m^2$ for the coupling.  

We can see 
that the NJL model is fundamental wherever we integrate out bosonic degrees of freedom but this result is more general than for this simple model as we are going to show below.

\subsection{$\phi^4$ model}
 
Now, let us consider a $\phi^4$ model extending the Yukawa model. 

A self-interacting bosonic field, when coupled to a set of fermionic fields by Yukawa couplings, has, as an effective low-energy field theory, a non-local Nambu-Jona-Lasinio model at the leading order, when higher powers of fermionic fields can be neglected.

We aim to use a current expansion as already proposed in the eighties by Cahill and Roberts \cite{Cahill:1985mh}. This will be also the track we will follow for QCD.

The Lagrangian has now the aspect
\begin{equation}
L=\frac{1}{2}\left[(\partial\phi)^2-\frac{\lambda}{2}\phi^4\right]+\sum_f\bar\psi_f\left(i\slashed\partial-g\phi-m_f\right)\psi_f+j\phi.
\end{equation}
We insert no mass term in the bosonic sector as this is obtained by the self-interaction \cite{Frasca:2009bc,Frasca:2013tma}. 
We assume $\phi=\phi[j]$. We can write \cite{Frasca:2013tma}
\begin{equation}
\phi[j]=\phi_0(x)+\int d^4x'\left.\frac{\delta\phi}{\delta j(x')}\right|_{j=0}j(x')+\frac{1}{2}\int d^4x'd^4x''\left.\frac{\delta^2\phi}{\delta j(x')\delta j(x'')}\right|_{j=0}j(x')j(x'')+O(j^3).
\end{equation}
We leave the fermion sector untouched unless for the dependence on $\phi$. We rewrite the above expansion as
\begin{equation}
\phi[j]=\phi_0(x)+\int d^4x'\Delta(x-x')j(x')+O(j^2)
\end{equation}
and we have recovered the previously discussed case. In fact, by substituting this expression into the Lagrangian, we obtain
\begin{equation}
L=L_0+\frac{1}{2}\int  d^4x' j(x)\Delta(x-x')j(x')+\sum_f\bar\psi_f\left(\slashed\partial-g\phi_0-g^2\int d^4x'\Delta(x-x')\bar\psi_f(x')\psi_f(x')-m_f\right)\psi_f+O(j^3)
\end{equation}
being
\begin{equation}
L_0=\frac{1}{2}\left[(\partial\phi_0)^2-\frac{\lambda}{2}\phi_0^4\right]
\end{equation}
the zeroth-order contribution arising from the exact solution of the quartic theory and provided that
\begin{equation}
\label{eq:Gphi4}
\partial^2\Delta+3\lambda\phi_0^2\Delta=\delta^4(x-x').
\end{equation}
The result is that, proceeding in this way, we again integrated out the bosonic degree of freedom. We observe from eq.~(\ref{eq:Gphi4}) that $\phi_0(x)$ cannot be taken to be zero. Then, by an ordering argument due to the structure of the scalar field propagator, obtained by solving eq.(\ref{eq:Gphi4}), we can consider the term $\phi_0(x)$, coupled to the fermionic sector, as a small perturbation and we are left again with a non-local NJL model. We will discuss this extensively for the case of QCD when the 1- and 2-point functions will be computed. In this case, the model is just a leading order approximation to the low-energy behavior of the theory as we are neglecting higher powers of fermion fields also arising from products of 2-point functions and higher order correlation functions of the scalar field.

From the above 
discussion
we see that the NJL model is obtainable only if we have a zeroth-order solution and we know the 2-point function. In quantum field theory this problem is completely equivalent to solve the Dyson-Schwinger hierarchy of equations at least for the 1- and 2-point functions. This has been accomplished for the scalar field (both with and without spontaneous symmetry breaking) in \cite{Frasca:2013tma,Frasca:2015wva,Frasca:2015yva} and for the Yang-Mills field \cite{Frasca:2015yva}. For these reasons, we can easily extend the above
result
to the case of QCD. In the next section, we will give explicitly the 1- and 2-point functions for both cases. Here we show that a similar 
result
%
holds also for QCD.

\subsection{Quantum Chromodynamics}

A Yang-Mills field, when minimally coupled to a set of fermionic fields, has as an effective field theory a non-local Nambu-Jona-Lasinio model at the leading order, when higher powers of fermionic fields can be neglected.

The argument runs essentially as in the previous 
computation
and is based again on the idea put forward by Cahill and Craig \cite{Cahill:1985mh}. We can write the QCD Lagrangian in the form
\begin{equation}
L=-\frac{1}{4}\operatorname{tr}\left(F^2\right)-\frac{1}{2\xi}(\partial\cdot A)^2-\sum_f\bar\psi_f\left(\slashed\partial-ig\frac{\lambda^a}{2}{\slashed A}^a\right)\psi_f
+j_\mu^aA^{a\mu}+\sum_f(\bar\eta_f\psi_f+\bar\psi_f\eta_f)+L_{gh}
\end{equation}
with $L_{gh}$ the ghost Lagrangian, provided $F_{\mu\nu}^a=\partial_\mu A_\nu^a-\partial_\nu A_\mu^a+gf^{abc}A_\mu^bA_\nu^c$, the parameter $\xi$ fixes the gauge choice and we are summing on all the flavors $f$. We have added the current terms like in the quartic scalar field. We are using the generator of the group $\lambda^a$ that satisfy the relation $[\lambda^a,\lambda^b]=if^{abc}\lambda^c$ with $f^{abc}$ the structure constants of the group. For SU(3), one has $a,b,c,\ldots=1\ldots 8$, for SU(N) the number of generators is $N^2-1$. 
Now, we can proceed like in the
previous
case 
using a Taylor series of functional derivatives. We will have
\begin{equation}
A_\nu^a[j]=A_\nu^a[0]+\int d^4x'\left.\frac{\delta A_\nu^a}{\delta j^{b\kappa}(x')}\right|_{j=0} j^{b\kappa}(x')
+\int d^4x'd^4x''\left.\frac{\delta^2 A_\nu^a}{\delta j^{b\kappa}(x')\delta j^{c\lambda}(x'')}\right|_{j=0} j^{b\kappa}(x')j^{c\lambda}(x'')+O(j^3)
\end{equation}
and we recognize the 1- and 2-point function $A_\nu^a[0]$ and $D_{\nu\kappa}^{ab}(x-x')=\left.\delta A_\nu^a/\delta j^{b\kappa}(x')\right|_{j=0}$.
Now, turning to the fermion sector of the QCD Lagrangian, one has for $j=0$ and substituting the above gluon field
\begin{equation}
L_f=\sum_f\bar\psi_f\left(i\slashed\partial-g\frac{\lambda^a}{2}A_\nu^a[0]-g^2\frac{\lambda^a}{2}\int d^4x' D_{\nu\kappa}^{ab}(x-x')\sum_{f'}\bar\psi_{f'}(x')\gamma^\kappa\frac{\lambda^b}{2}\psi_{f'}(x')\right)\psi_f+\ldots.
\end{equation}
Dots imply higher order powers of the fermion fields. We get a quadratic functional for the gluon field given by
\begin{equation}
L_g=\frac{1}{2}\int d^4x'j^{a\mu}(x)D_{\mu\nu}^{ab}(x-x'')j^{b\nu}(x'')+O(j^3).
\end{equation}
If we choose the Landau gauge, the 2-point function takes the form
\begin{equation}
D_{\mu\nu}^{ab}(x-x')=\delta^{ab}\left(\eta_{\mu\nu}-\frac{\partial_\mu\partial_\nu}{\partial^2}\right)\Delta(x-x')
\end{equation}
that permits us to simplify the fermion Lagrangian as
\begin{equation}
L_f=-\sum_f\bar\psi_f\left(\slashed\partial-ig\frac{\lambda^a}{2}A_\nu^a[0]\right)\psi_f-g^2\sum_f\sum_{f'}\int d^4x'\Delta(x-x')\bar\psi_f(x)\gamma_\kappa\frac{\lambda^a}{2}
\bar\psi_{f'}(x')\gamma^\kappa\frac{\lambda^a}{2}\psi_{f'}(x')\psi_f(x)+\ldots.
\end{equation}
We again integrated out the bosonic degrees of freedom and we can see that we have recovered again a non-local NJL model with a quartic self-interacting fermion field emerging directly from QCD. 

It is not difficult to see that, to quantize the theory, we need the n-point functions obtained from the Dyson-Schwinger equations. Therefore, our aim in the following section will be to give explicitly these 1- and 2-point functions to evaluate properly this low-energy limit of QCD, choosing the Landau gauge. Landau gauge grants the knowledge of exact solutions for the 1-point function and a simple expression for the 2-point function
. We will also see how an ordering argument dumps down the coupling between the quark fields and 1-point function of the Yang-Mills field.

%
\section{1- and 2-point functions for Yang-Mills theory}
\label{sec1}


We start by analyzing the case of the self-interacting scalar field theory with the equation of motion
\begin{equation}
\label{eq:cphi}
   \Box\phi+\lambda\phi^3=j.
\end{equation}
The homogeneous equation with $j=0$ admits the exact solution \cite{Frasca:2009bc}
\begin{equation}
\label{eq:phi0}
\phi_0(x) = \mu\left(2/\lambda\right)^\frac{1}{4}{\rm sn}(p\cdot x+\theta,i)
\end{equation}
being {\rm sn} an elliptic Jacobi function and $\mu$ and $\theta$ two integration constants. This holds provided the following dispersion relation is satisfied
\begin{equation}
p^2=\mu^2\sqrt{\lambda/2}.
\end{equation}
This represents a free massive solution notwithstanding we started from a massless theory. Mass arises from the nonlinearity in the equation of motion provided $\lambda$ stays finite rather than going to zero. Indeed, standard perturbation theory just fails to recover it. Moving to quantum field theory, we have to evaluate the n-point functions obtained by Dyson-Schwinger equations. Limiting our interest to 1- and 2-point functions, we will write \cite{Frasca:2015yva}
\begin{eqnarray}
&&\partial^2 G_1(x)+\lambda\left([G_1(x)]^3+3G_2(0)G_1(x)+G_3(0,0)\right)=0 \nonumber \\
   &&\partial^2G_2(x-y)+\lambda\left(3[G_1(x)]^2G_2(x-y)+3G_2(0)G_2(x-y)\right. \nonumber \\
	&&\left.+3G_3(0,x-y)G_1(x)+G_4(0,0,x-y)\right)=\delta^4(x-y).
\end{eqnarray}
These equations can be solved exactly by noting that we have a mass correction $3\lambda G_2(0)$, that we assume small after renormalization so that we can work with eq.(\ref{eq:phi0}) (see \cite{Frasca:2017slg}). Then, the contributions from higher-order n-point functions are taken to be 0 \cite{Frasca:2015yva}. 
We make an approximation at this stage by neglecting mass corrections. This makes the propagator simpler to work with. Anyway, exact expression is given in \cite{Frasca:2017slg} where is also shown an exceedingly good agreement with lattice data for the theory spectrum, both in 3 and 4 dimensions. 
Granted these points, we will
work with
the propagator \cite{Frasca:2009bc}
\begin{equation}
\label{eq:green}
   \Delta(p)=\sum_{n=0}^\infty\frac{B_n}{p^2-m_n^2+i\epsilon}
\end{equation}
being
\begin{equation}
   B_n=(2n+1)^2\frac{\pi^3}{4K^3(-1)}\frac{e^{-(n+\frac{1}{2})\pi}}{1+e^{-(2n+1)\pi}}
\end{equation}
and 
\begin{equation}
\label{eq:spec}
   m_n=(2n+1)(\pi/2K(-1))\left(\lambda/2\right)^{\frac{1}{4}}\mu
\end{equation} 
being $K(-1)=1.3111028777\ldots$ an elliptic integral. This holds provided one fixes the phase $\theta$ in the exact solution to $\theta_m=(4m+1)K(-1)$ to preserve translation invariance in the propagating degrees of freedom. It identifies an infinite set of scalar field theories with a trivial infrared fixed point in quantum field theory. This propagator gives explicitly the nNJL model obtained 
from $\phi^4$ model
in a general form, solving completely the low-energy limit for this theory. The local limit is obtained by taking $p\rightarrow 0$ in eq.(\ref{eq:green}) that provides
\begin{equation}
   \Delta(x-x')=-\sum_{n=0}^\infty\frac{B_n}{m_n^2}\delta^4(x-x').
\end{equation}

This can be immediately applied to Yang-Mills theories as we have exact solutions also for this case \cite{Frasca:2015yva}. We report here the Dyson-Schwinger equations for 1- and 2-point functions for the Yang-Mills theory, in the Landau gauge, as computed in \cite{Frasca:2015yva} 
\begin{eqnarray}
G_{1\mu}^a(x)&=&\eta_\mu^a\chi(x) \nonumber \\
G_{2\mu\nu}^{ab}(x-y)&=&\delta_{ab}\left(g_{\mu\nu}-\frac{p_\mu p_\nu}{p^2}\right)\Delta(x-y)
\end{eqnarray} 
Note that the longitudinal part of the propagator is 0 because is proportional to gauge fixing parameter that is 0 in the Landau gauge. This is no more true at finite temperature. Here the ghost field decoupled and $\phi(x)$ and $\Delta(x-y)$ satisfying the Dyson-Schwinger equations \cite{Frasca:2015yva}
\begin{eqnarray}
\label{eq:DS-YM}
\partial^2\chi(x)+2Ng^2\delta\mu^2\chi(x)+Ng^2\chi^3(x)&=&0  \nonumber \\
\partial^2\Delta(x-y)+2Ng^2\delta\mu^2\Delta(x-y)+3Ng^2\phi^2(x-y)\Delta(x-y)&=&\delta^4(x-y) \nonumber \\
\partial^2 P^{am}_2(x-y)&=&\delta_{am}\delta^4(x-y).
\end{eqnarray}
$P^{am}_2(x-y)$ is the ghost propagator that appears completely decoupled,
$\chi(x)$ and $\Delta(x-y)$ coincide exactly with eqs.(\ref{eq:phi0}) and (\ref{eq:green}), provided we substitute $\lambda\rightarrow Ng^2$, and taking into account that the evaluation of the $\delta\mu^2$ contribution has been given in \cite{Frasca:2017slg}.This, after renormalization, is proven to yield just a small correction to the spectrum of the theory given by eq.(\ref{eq:spec}). Therefore, we can safely work neglecting it and we will work with the approximate expression given in eq.~(\ref{eq:green}). These solutions confirm that also Yang-Mills theories seem to share a trivial infrared fixed point. This is supported by lattice studies of the running coupling \cite{Bogolubsky:2009dc} from lattice at $64^4$ and $80^4$ with $\beta=5.7$ where the running coupling is seen to go to zero as momenta lower. A similar result was obtained in \cite{Boucaud:2002fx}. This latter computation shows a perfect consistency with an instanton liquid model in agreement with our scenario as we will see below.

\begin{figure}[H]
  \includegraphics{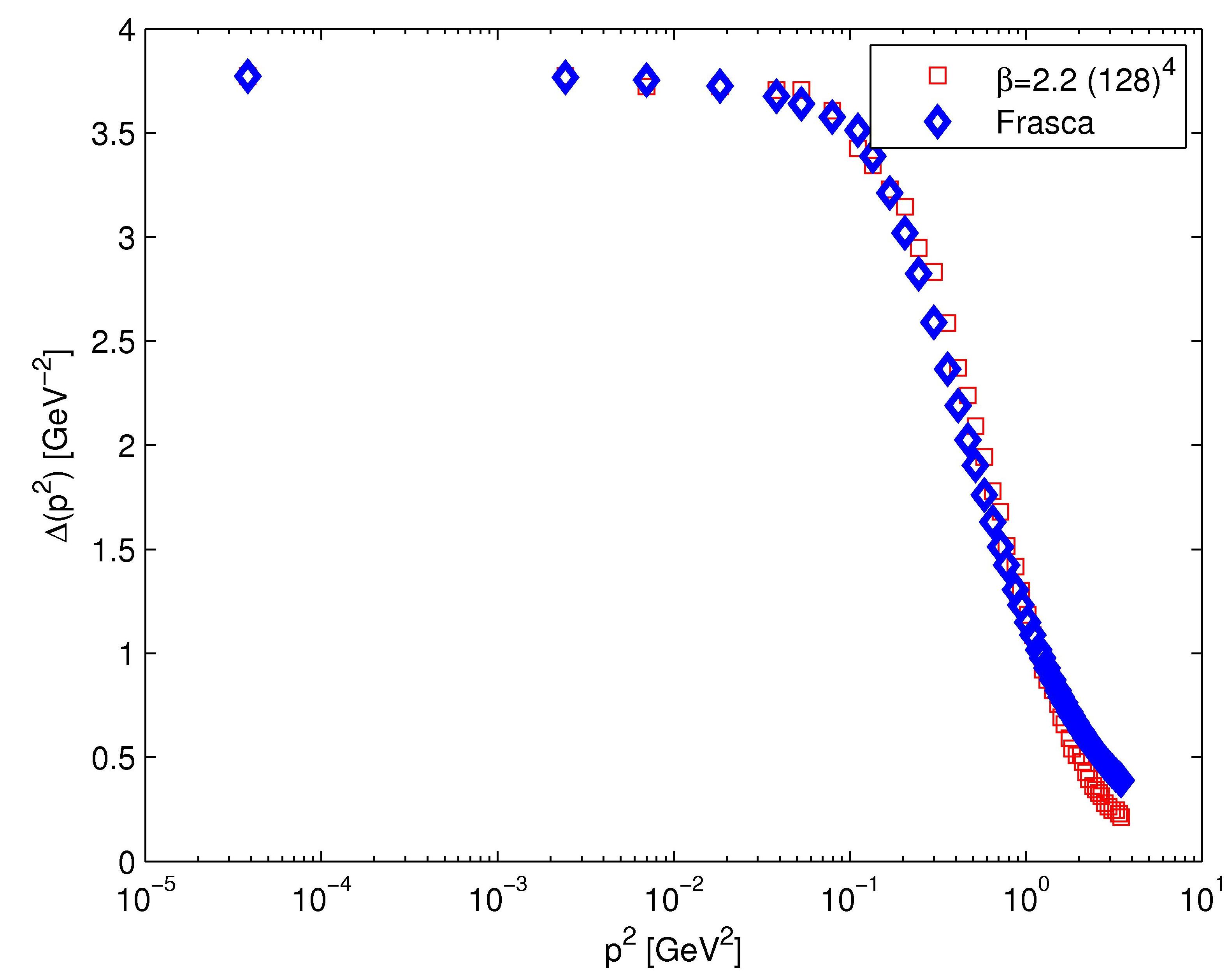}
  \caption{Comparison of our propagators with the lattice data for SU(2) given in \cite{Cucchieri:2007md} for $(128)^4$ points.\label{fig:cuccmend}}
\end{figure}

\begin{figure}[H]
  \includegraphics{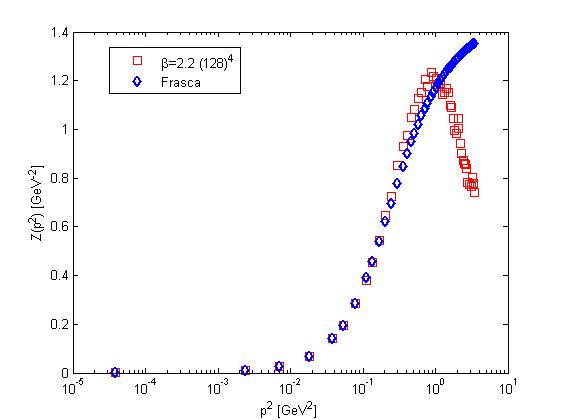}
  \caption{Same as in Fig.~\ref{fig:cuccmend} but for the dressing function of the gluon propagator. \label{fig:dress}}
\end{figure}

Then, the generating functional for the scalar field 
at the leading order, 
is just a Gaussian generating functional with the propagator given by eq.~(\ref{eq:green}). Next-to-leading orders can also be computed \cite{Frasca:2013tma}. Similarly, turning our attention to the Yang-Mills generating functional, we realize that it also takes the simple Gaussian form, at the leading order,
\begin{equation}
     Z_0[j]=N\exp\left[\frac{i}{2}\int d^4x'd^4x''j^{a\mu}(x')D_{\mu\nu}^{ab}(x'-x'')j^{b\nu}(x'')\right],
\end{equation}
being $D_{\mu\nu}^{ab}(x-x')$ has been just obtained from the Dyson-Schwinger equations (\ref{eq:DS-YM}) and, as already stated, is the same of eq.~(\ref{eq:green}) provided we exchange $\lambda\rightarrow Ng^2$.
This propagator represents a sum of propagators of a free theory and a mass spectrum of glue excitations identical to that of a harmonic oscillator. Ghost field just decouples
yielding a free massless propagator. All these properties of the quantum Yang-Mills field correspond to the so-called ``decoupling solution'' \cite{Aguilar:2004sw,Boucaud:2006if,Frasca:2007uz} (see also \cite{Weber:2011nw} for a discussion) that also implies a decoupled ghost propagator as we get. This kind of gluon propagator is the one recovered in lattice computations \cite{Bogolubsky:2007ud,Cucchieri:2007md,Oliveira:2007px}. We yield a comparison in Fig.\ref{fig:cuccmend}
 where we consider just a single fitting parameter given by $m_0=(\pi/2K(-1))\left(Ng^2/2\right)^{\frac{1}{4}}\mu$ that we take to be $0.436172183\ GeV$, very near the string tension, generally taken to be $0.44\ GeV$.
For the sake of completeness, we report here also the dressing function of the gluon propagator, defined as $Z(p^2)=p^2\Delta(p)$, for our propagator and the lattice one. At higher momenta we do not expect a complete agreement as we are using an approximate solution, neglecting the effects due to the running coupling. In Fig.\ref{fig:dress} we see the peak in the lattice data at around 1 GeV that is very near to the plateau formation in our approximation.
%
%
The agreement is exceedingly good at low energies as expected. The exact solutions discussed in this section should grant a completion of the low-energy limit 
providing a well-defined non-local NJL model that we will discuss in the next section.

\section{Low-energy limit of QCD}
\label{sec10}

\begin{figure}[H]
  \includegraphics{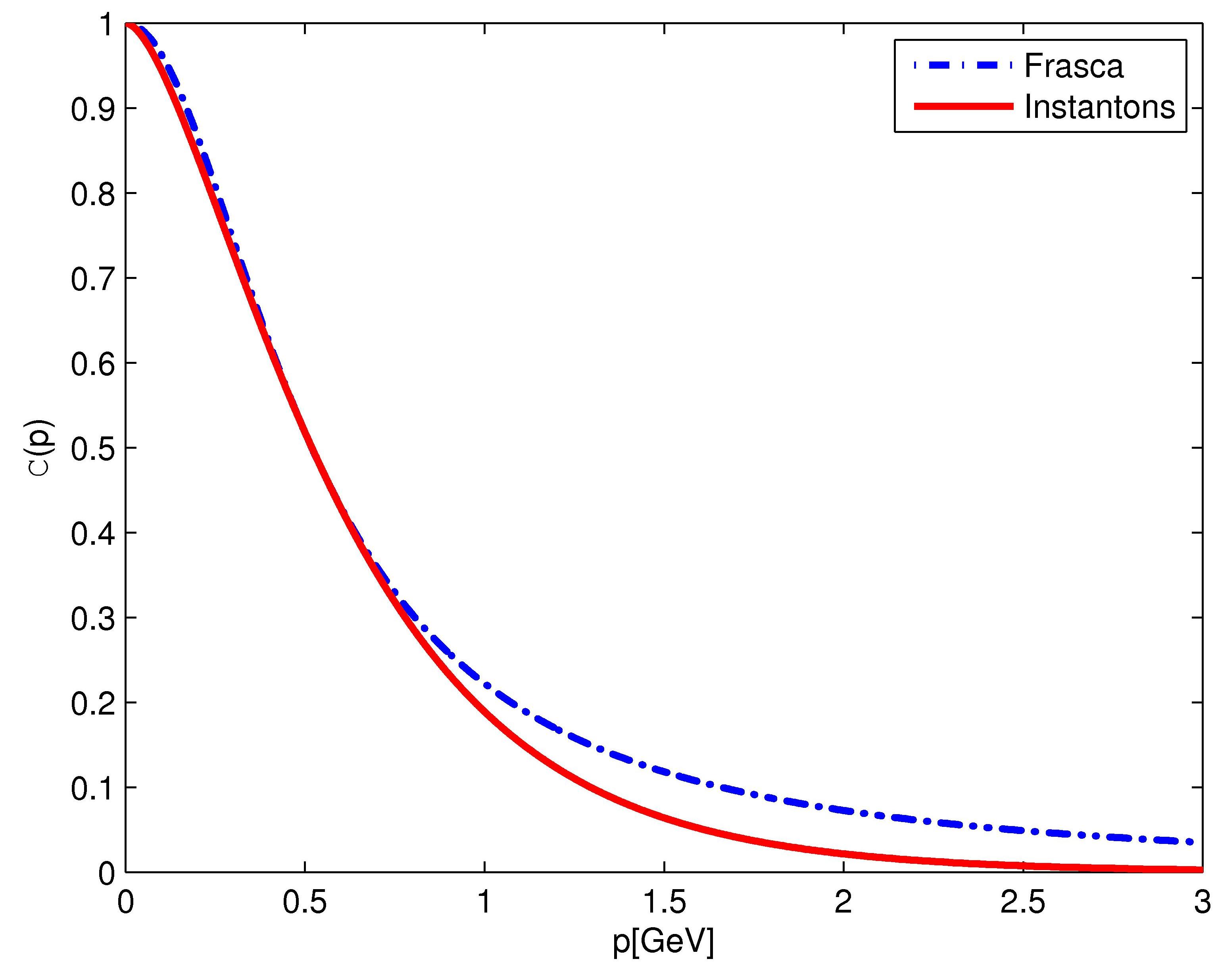}
  \caption{Comparison of our form factor with that provided in \cite{Schafer:1996wv} 
  for $\sqrt{\tilde\sigma}=0.417\ GeV$ and $d^{-1}=0.58\ GeV$.\label{fig:ff}}
\end{figure}

The results discussed in the previous sections permit to derive the low-energy limit of QCD and this, as said, coincides with a non-local Nambu-Jona-Lasinio model at the leading order \cite{Frasca:2011bd,Frasca:2012eq,Frasca:2012iv,Frasca:2013kka}. Here, we exploit this derivation step by step to be as clearer as possible. The contribution coming from the Gaussian contribution of the gluon field can be stated as
\begin{equation}
S_g=\int d^4x'd^4x''j^{a\mu}(x')D_{\mu\nu}^{ab}(x'-x'')j^{b\nu}(x'')=\sum_{n=0}^\infty B_n\int\frac{d^4p}{(2\pi)^4}j^{a\mu}(p)\frac{1}{p^2-m_n^2+i\epsilon}j^{a}_\mu(-p).
\end{equation}
This action represents the sum of infinite scalar fields and is weighted by exponentially damped coefficients $B_n$. So, we assume we can neglect all the contributions coming from the excitations of the gluon filed at $n>0$. Then, taking just the leading contribution at $n=0$, we approximate  
\begin{equation}
      S_g=\int d^4x\left[\frac{1}{2}(\partial\sigma)^2-\frac{1}{2}m_0^2\sigma^2\right]+S_q
\end{equation}
where we named $\sigma$ this field arising from the gluon propagator in the Gaussian generating functional of the Yang-Mills action with all the higher order excited state in the superimposed harmonic oscillator spectrum being exponential neglected. This is the contribution arising from the mass gap of the theory, being $m_0=(\pi/2K(i))\sqrt{\tilde\sigma}$ and will add to the $\sigma$ field coming out from the fermion action. Here and in the following we will assume $\tilde\sigma$ for the string tension (normally estimated to be about $(440\ MeV)^2$). So, quark fields yield
\begin{eqnarray}
\label{eq:njl}
      S_q&=&\sum_q\int d^4x\bar q(x)\left[i{\slashed\partial}-m_q-g\sqrt{\frac{B_0}{3(N_c^2-1)}}
      \eta_\mu^a\gamma^\mu\frac{\lambda^a}{2}\sigma(x)\right]q(x) \\  
     &-&g^2\int d^4x'\Delta(x-x')\sum_q\sum_{q'}\bar q(x)\frac{\lambda^a}{2}\gamma^\mu\bar q'(x')\frac{\lambda^a}{2}\gamma_\mu q'(x')q(x)
      +O\left(\frac{1}{\sqrt{N}g}\right)+O\left(j^3\right). \nonumber
\end{eqnarray}
with the scale factor $\sqrt{\frac{B_0}{3(N_c^2-1)}}$ arising both from the definition of the field $\sigma$, yielding the $B_0$ contribution, and the normalization condition in the Landau gauge $\eta_\mu^a\eta_\mu^a=\delta_{ab}(g_{\mu\nu}-p_\mu p_\nu/p^2)$, arising from the 1-point function. This damps out the coupling between the 1-point solution and the quark fields by two magnitude orders with respect to the other terms, relegating this to a very small perturbation. Then, our non-local Nambu-Jona-Lasinio model coincides with that presented in \cite{Hell:2008cc}, directly from QCD, provided the form factor is
\begin{equation}
\label{eq:Gp}
      {\cal G}(p)=-\frac{1}{2}g^2\Delta(p)=-\frac{1}{2}g^2\sum_{n=0}^\infty\frac{B_n}{p^2-(2n+1)^2(\pi/2K(i))^2\tilde\sigma+i\epsilon}
      =\frac{G}{2}{\cal C}(p)
\end{equation}
being $B_n$ obtained from eq.(\ref{eq:green}), ${\cal C}(0)=1$ and $2{\cal G}(0)=G$ the standard Nambu-Jona-Lasinio coupling, fixing in this way the value of $G$ through the gluon propagator. In Fig.~\ref{fig:ff}, we compare this form factor both with the one from an instanton liquid \cite{Schafer:1996wv} that is
\begin{equation}
\mathcal{C}_I(p)=p^2 d^2\left\{\pi \dfrac{d}{d\xi}\big[I_0(\xi)K_0(\xi)-I_1(\xi)K_1(\xi)\big]\right\}^2\qquad\text{with } \xi=\frac{|p| d}{2} 
\end{equation}
being $I_n$ and $K_n$ Bessel functions. In the following we normalize this function to be 1 at zero momenta dividing it by ${\cal C}_I(0)$.
The result is strikingly good for our form factor showing how consistently our technique represents Yang-Mills theory through instantons. In the low-energy limit recovers a non-local Nambu-Jona-Lasinio model and maintains the defects of this approximation as a non-confining behavior. Higher order corrections can grant to recover this property of the theory. With our approach these can be computed. Anyhow, it should be noted a slower decay to infinity of the blue curve in Fig.~\ref{fig:ff} but, extending the momentum scale, it seen to run to zero properly as expected. So, there is no problem in the high-energy limit.

So, finally we write down the NJL action we will use in the following as was obtained from QCD
\begin{eqnarray}
\label{eq:njlok}
      S_q&=&\sum_q\int d^4x\bar q(x)\left[i{\slashed\partial}-m_q\right]q(x) \nonumber \\  
     &+&\int d^4x\int d^4x'{\cal G}(x-x')\sum_q\sum_{q'}\bar q(x)\frac{\lambda^a}{2}\gamma^\mu\bar q'(x')
		\frac{\lambda^a}{2}\gamma_\mu q'(x')q(x).
\end{eqnarray}
This can be bosonized in a standard way \cite{Ebert:1994mf,Bernard:1993rz} giving the effective field theory. One introduces the field $\sigma(x)=G\bar q(x)q(x)$ and ${\bm\pi}(x)=G\bar q(x)\gamma^5{\bm\tau}q(x)$ and this will yield, after a Fierz rearrangement and considering a two flavor QCD,
\begin{eqnarray}
\label{eq:LB}
   S_{NJL}&=&\int d^4x\frac{1}{2}(\partial\sigma)^2-\frac{1}{2}m_0^2\sigma^2\nonumber \\
   &+&\int d^4x\sum_{q=\{u,d\}}\bar q (i{\slashed \partial} - g(\sigma +  i\gamma^5 {\bm\pi}\cdot{\bm\tau})q) \nonumber \\
	&-&\frac{1}{2}\int d^4x\int d^4x'{\cal C}(x-x')\left({\sigma^2}(x')+{\bm\pi}(x')\cdot{\bm\pi}(x')\right).
\end{eqnarray}
being ${\bm\tau}$ SU(2) Pauli matrices and neglecting quark masses. This appears as a well-known quark-meson model and so, we can add the chemical chiral potential as in \cite{Ruggieri:2011xc}
\begin{equation}
   S_{NJLc} = S_{NJL} + \int d^4x\sum_{q=\{u,d\}}\mu_5\bar q\gamma^0\gamma^5q.
\end{equation}
Finally, we will perform all the computations at finite temperature.

\section{Critical temperature}
\label{sec2}

The potential has the form \cite{Hell:2008cc}
\begin{eqnarray}
    V(\sigma,{\bm\pi})&=&-i{\rm Trln}\left[1-(i {\slashed \partial} - \hat m -\gamma_0\gamma_5\mu_5)^{-1}
    (\sigma +  i\gamma^5 {\bm\pi}\cdot{\bm\tau})\right] \nonumber \\
    &+&\int d^4x\left[\frac{1}{2}(G^{-1}+m_0^2)\sigma^2+\frac{1}{2G}{\bm\pi}\cdot{\bm\pi}\right]
\end{eqnarray}
that yields the gap equation \cite{Hell:2008cc}
\begin{equation}
\label{eq:v0}
    v=\frac{4N_cN_f}{m_0^2+1/G}\beta^{-1}\sum_{k=-\infty}^\infty\sum_{s=\pm 1}\int\frac{d^3p}{(2\pi)^3}{\cal C}(\omega_k,|{\bm p}|s+\mu_5)\frac{M(\omega_k,|{\bm p}|s+\mu_5)}
    {\omega_k^2+(|{\bm p}|s+\mu_5)^2+M^2(\omega_k,|{\bm p}|s+\mu_5)},
\end{equation}
having set
\begin{equation}
    M(|{\bm p}|s+\mu_5)={\cal C}(|{\bm p}|s+\mu_5)v,
\end{equation}
being $v$ the vacuum expectation value of the $\sigma$ field. We have introduce a sum on the Matsubara frequencies $\omega_k=(2k+1)T$.
Here ${\cal C}(p)$ is given by $\frac{G}{2}{\cal C}(p)={\cal G}(p)$ using eq.(\ref{eq:Gp}) but moving to Euclidean. The restoration of chiral symmetry is given at $v=0$ and so, we have to solve
\begin{equation}
\label{eq:v1}
    1=\frac{4N_cN_f}{m_0^2+1/G}\beta^{-1}\frac{g^4}{G^2}\sum_{k=-\infty}^\infty\sum_{s=\pm 1}
		\int\frac{d^3p}{(2\pi)^3}{\cal C}^2(\omega_k,|{\bm p}|s+\mu_5)\frac{1}
    {\omega_k^2+(|{\bm p}|s+\mu_5)^2}
\end{equation}
to obtain the critical temperature as a function of $\mu_5$. We consider just one term in the form factor (\ref{eq:Gp}). This is so because we want to be consistent with the NJL action just obtained, noting that higher excitations are exponentially damped. Then, we will have
\begin{equation}
    {\cal C}(p)=\frac{g^2}{G}\frac{B_0}{p^2+m_0^2}
\end{equation}
having moved to Euclidean and being $m_0=(\pi/2K(i))\sqrt{\tilde\sigma}$ the mass gap. We take $Z=g^2B_0/G$ and then
\begin{equation}
\label{eq:v2}
    1=\frac{4N_cN_f}{m_0^2+1/G}\beta^{-1}\sum_{k=-\infty}^\infty\sum_{s=\pm 1}
		\int\frac{d^3p}{(2\pi)^3}\frac{Z^2}{(\omega_k^2+(|{\bm p}|s+\mu_5)^2+m_0^2)^2}\frac{1}
    {\omega_k^2+(|{\bm p}|s+\mu_5)^2}.
\end{equation}
Matsubara sum can be performed analytically giving
\begin{eqnarray}
    {\cal I}_{p,s}&=&\beta\frac{\pi}{2m_0^4||{\bm p}|s+\mu_5|}\tanh\left(\frac{\pi}{2}\beta||{\bm p}|s+\mu_5|\right) \nonumber \\
		&&+\beta^2\frac{\pi^2}{8((|{\bm p}|s+\mu_5)^2+m_0^2)m_0^2} \nonumber \\
		&&-\beta^2\frac{\pi^2}{8((|{\bm p}|s+\mu_5)^2+m_0^2)m_0^2}
		\tanh^2\left(\frac{\pi}{2}\beta\sqrt{(|{\bm p}|s+\mu_5)^2+m_0^2}\right) \nonumber \\
		&&-\beta\frac{\pi}{4}\frac{2(|{\bm p}|s+\mu_5)^2+3m_0^2}{((|{\bm p}|s+\mu_5)^2+m_0^2)^\frac{3}{2}m_0^4}
		\tanh\left(\frac{\pi}{2}\beta\sqrt{(|{\bm p}|s+\mu_5)^2+m_0^2}\right).
\end{eqnarray}
In order to get an understanding, we try to solve the gap equation with $\mu_5=0$ and then, we restate it into the equation. We will have
\begin{eqnarray}
    {\cal I}_{p,1}&=&\beta\frac{\pi}{2m_0^4p}\tanh\left(\frac{\pi}{2}\beta p\right) \nonumber \\
		&&+\beta^2\frac{\pi^2}{8(p^2+m_0^2)m_0^2} \nonumber \\
		&&-\beta^2\frac{\pi^2}{8(p^2+m_0^2)m_0^2}\tanh^2\left(\frac{\pi}{2}\beta\sqrt{p^2+m_0^2}\right) \nonumber \\
		&&-\beta\frac{\pi}{4}\frac{2p^2+3m_0^2}{(p^2+m_0^2)^\frac{3}{2}m_0^4}
		\tanh\left(\frac{\pi}{2}\beta\sqrt{p^2+m_0^2}\right)
\end{eqnarray}
that yields for $\beta\rightarrow 0$, after integration on momenta,
\begin{equation}
    \int\frac{d^3p}{(2\pi)^3}{\cal I}_{p,1}\stackrel{\beta\rightarrow 0}{=}\beta^2\frac{\pi^2}{8m_0^2}\Lambda
\end{equation}
being $\Lambda$ a needed cut-off to regularize divergent integrals. This cut-off must be chosen so that the product $\beta\Lambda$  is kept constant while $\Lambda$ runs to infinity. Then,
\begin{equation}
    T_c=Z^2\frac{N_cN_f}{m_0^2+1/G}\frac{g^4}{G^2}\frac{\pi^2}{2m_0^2}\Lambda.
\end{equation}
We see that, in this case, the gap equation admits always a solution whatever is the coupling $g$ and chiral symmetry is broken. As expected, temperature runs to infinity as the cut-off itself.

When $\mu_5$ is turned on, one has, taking the limit $\beta\rightarrow 0$,
\begin{eqnarray}
    {\cal I}_{p,s}&=&\beta^2\frac{\pi^2}{4m_0^4} \nonumber \\
		&&+\beta^2\frac{\pi^2}{8((|{\bm p}|s+\mu_5)^2+m_0^2)m_0^2} \nonumber \\
		&&-\beta^2\frac{\pi^2}{8}\frac{2(|{\bm p}|s+\mu_5)^2+3m_0^2}{((|{\bm p}|s+\mu_5)^2+m_0^2)m_0^4}.
\end{eqnarray}
and, after integration on momenta, we get
\begin{equation}
		\sum_{s=\pm 1}\int\frac{d^3p}{(2\pi)^3}{\cal I}_{p,s}=\beta^2\frac{1}{4m_0^2}\Lambda+\beta^2\frac{\pi}{8m_0^3}(\mu_5^2-m_0^2).
\end{equation} 
This yields
\begin{equation}
     T_c=Z^2\frac{N_cN_f}{m_0^2+1/G}\frac{g^4}{G^2}\left(\frac{1}{4m_0^2}\Lambda+\frac{\pi}{8m_0^3}(\mu_5^2-m_0^2)\right).
\end{equation}
This is the main result of the paper showing that the critical temperature increases with the chiral chemical potential in agreement with Ruggieri and Peng~\cite{Ruggieri:2016cbq}, with lattice results~ \cite{Braguta:2015owi,Braguta:2015zta} and solution of Dyson-Schwinger equations~\cite{Xu:2015vna}. 

It is interesting to note the dependence on the mass gap $m_0$. This equation seems to imply that $|\mu_5|\ge m_0$ but for the all practical purposes, the cut-off $\Lambda$ is large enough with respect to the mass gap to grant always a physical value for $T_c$ also when $|\mu_5|<m_0$.

\section{Conclusions}
\label{sec3}

Using recent studies on lattice, we were able to derive the low-energy limit of QCD. The result is given by a nonlocal NJL model that is amenable to analytical computations. In this way, we are able to conclude that the critical temperature for chiral symmetry breaking in QCD increases as the square of the chiral chemical potential in agreement with recent lattice studies. This result supports the conclusions presented in a recent work \cite{Ruggieri:2016cbq} supporting a preferential choice of a renormalization scheme.
 
\begin{acknowledgments}
I would like to thank Marco Ruggieri for the insightful discussions about this matter that motivated this paper.
\end{acknowledgments}

\end{document}